# Discovery of Volcanic Activity on Io

## A Historical Review

Linda A. Morabito[1]

[1]Department of Astronomy, Victor Valley College, Victorville, CA 92395

linda.morabito@vvc.edu

Borrowing from the words of William Herschel about his discovery of Uranus:

'It has generally been supposed it was a lucky accident that brought the volcanic plume to my view. This is an evident mistake. In the regular manner I examined every Voyager 1 optical navigation frame. It was that day the volcanic plume's turn to be discovered."
– Linda Morabito

Abstract

In the 2 March 1979 issue of Science **203** S. J. Peale, P. Cassen and R. T. Reynolds published their paper "Melting of Io by tidal dissipation" indicating "the dissipation of tidal energy in Jupiter's moon Io is likely to have melted a major fraction of the mass." The conclusion of their paper was that "consequences of a largely molten interior may be evident in pictures of Io's surface returned by Voyager 1." Just three days after that, the Voyager 1 spacecraft would pass within 0.3 Jupiter radii of Io. The Jet Propulsion Laboratory navigation team's orbit estimation program as well as the team members themselves performed flawlessly. In regards to the optical navigation component image extraction of satellite centers in Voyager pictures taken for optical navigation at Jupiter rms post fit residuals were less than 0.25 pixels. The cognizant engineer of the Optical Navigation Image Processing system was astronomer Linda Morabito. Four days after the Voyager 1 encounter with Jupiter, after performing image processing on a picture of Io taken by the spacecraft the day before, something anomalous emerged off the limb of Io. This historical review written by the discoverer recounts her minute-by-minute quest to identify what was a volcanic plume, the first evidence of active volcanism seen beyond Earth. Many ingredients of this account reflect historic themes in the process of scientific discovery.

1. Introduction

This account of the discovery of the active volcanism on Jupiter's moon Io is taken from an 11 page single spaced typed letter written by Linda Morabito on a typewriter at her home in Pasadena, California on March 15, 1979 to the then chairman of the department of astronomy at the University of Southern California, the late Professor Emeritus Gibson Reaves. Gibson Reaves placed the letter into the archives at USC within days of his having received it, where it remains. Gibson Reaves, who retired from his teaching position at USC in the fall of 1994 after a forty-two year career there was distinguished for his pioneering studies in extragalactic astronomy in regards to dwarf galaxies and his work in the history of astronomy. He is credited with the discovery of 5 supernovae at the 48-inch Schmidt telescope on Palomar Mountain. Professor Reaves was also a recipient of the USC Associates Award for teaching excellence in 1974, the same year in which USC Astronomy major Linda Morabito had received her undergraduate degree. When Linda Morabito contacted Gibson Reaves by telephone on the evening of Sunday, March 11, 1979 to inform him about her discovery, Gibson instructed her:

"Linda, write everything down."

By mid-day the next day, the news about the discovery of currently active extraterrestrial volcanism on Io had gone around the world.

There are countless articles that have been written and are still being written about the Voyager 1 and Voyager 2 spacecraft mission. We shall concentrate on personal reflections about the real-time process of scientific discovery, through the eyes of a young scientist working as a mission navigator on the mission. My reflections are based on the aforementioned historical document given to my mentor and life-long friend Gibson Reaves. First, however, is the legacy of a mission that showed the remarkable accuracy and success it did in science as well as engineering and defined a Golden Age in space exploration.

2. Voyager's Legacy

The two Voyager spacecraft, Voyagers 1 and 2, that launched in August and September of 1977 respectively and in reverse order of their arrival at Jupiter rewrote the textbooks on our knowledge of the outer solar system. Dr. Edward Stone, the Voyager project scientist has described the Voyager mission as:

"The one mission that went to more new worlds than any mission before it ever had, or perhaps any mission ever will."

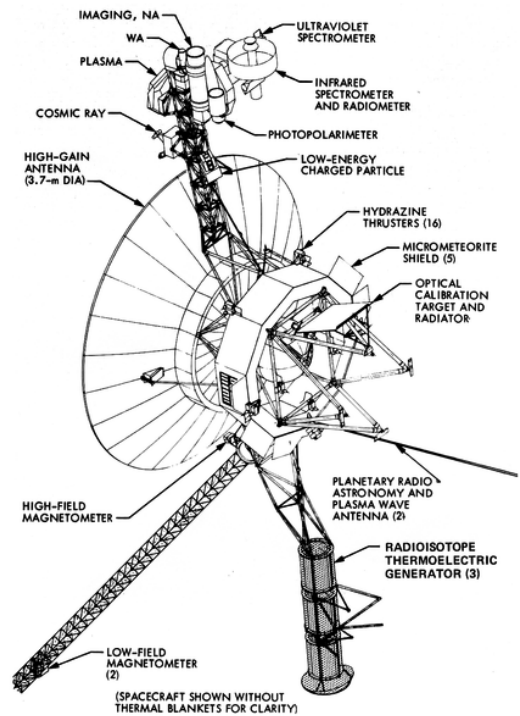

After a journey of 546 days and more than 1 billion miles Voyager 1 returned some 15,000 high resolution pictures of Jupiter and its moons for science (Campbell 1983). However, only 93 pictures were planned and taken by Voyager 1 for optical navigation for its Jupiter encounter.

The narrow angle imaging science instrument (labeled "IMAGING, NA" in Figure 1) was used for the navigation pictures that included at least one satellite of Jupiter against a star background. The instrument resolution was 10 μradians. The exposure time of these optical navigation images was 960 msec; long enough to ensure detection of stars up to 9.5 effective visual magnitude. At this exposure.

Figure 1. Drawing of Voyager Spacecraft
Credit NASA

three out of four of the Galilean moons of Jupiter were between two and four times overexposed.

Linda Morabito was appointed System Engineer for Optical Navigation Image Processing in the Navigation Systems Section (314) at JPL effective June 19, 1978. Her appointment was approved by Dr. J. F. Jordan, Manager of the Section and by Dr. C. R. Gates who was the Systems Division 31 Manager. Her appointment was agreed to Norri Sirri, MCCC Mark III Program Manager.

Just prior to her appointment, Linda and her co-worker Homa Taraji had visited Lick Observatory to select stars in photographic plates taken there. The ultimate goal was to create a star catalog with up-to-date equatorial coordinates of reference stars in a band of the sky against which cameras of the Voyager spacecraft might be aligned for observations of the satellites of Jupiter during the flyby (Klemola 1978). They selected nearly 5000 stars for measuring using the Lick Gaertner automatic measuring machine over two long two-day sessions.

Additionally Linda had done a pre-flight calibration of the distortion expected in pictures taken by potential Voyager flight cameras. She had actually developed the model that would be used to determine distortion in pictures that would be processed by the ONIPS MODCOMP IV (the Optical Navigation Image Processing System minicomputer) using dark marks etched in the camera for this purpose, called reseau marks.

Linda had researched formulations and design, and implemented a computer program for the prediction of good opportunities for Optical Navigation pictures based on satellite (moon) to satellite and satellite to star proximity, since all of the pictures for Optical Navigation required at least one of the moons of Jupiter against a star background in each picture. Before the Voyager spacecrafts had even launched, she had created a pre-launch picture schedule needed for the Optical Navigation. Additionally, she had become familiar with conventions and accuracies of existing major star catalogs in the development of a Star Catalog Software System used to generate small star files necessary for the on-board navigation of Voyager. Linda had produced four Internal Engineering Memorandum in 1975 at JPL to document this work. Previously she had served on the Viking mission to Mars flight team, and had been selected as a co - guest investigator on the Viking extended mission along with Tom Duxbury. Linda was given the directive of conducting the Phobos/Deimos Shadow Experiment to determine the locations of the Viking Landers on the surface of Mars using imaging data of the shadow of Phobos/Deimos respective passage over the each lander (Duxbury 1976).

On 6 September 1978, Linda signed off on the approved ONIPS User's Guide as ONIPS Cognizant Engineer, along with J. P. McDanell, Outer Planet Navigation Group Supervisor, and Norri Sirri. The center-finding software for the image extraction in pictures planned for Optical Navigation had been finalized under Linda's cognizance. Additionally, Linda had successfully mathematically modeled geometric blooming or distortion expected in the overexposed satellite images due to vidicon "beam bending." Left unaccounted for, beam-bending distortion would have yielded pixel-level center-finding errors. Linda's operation of the Optical Navigation Image Processing System for the navigation of the Voyager spacecraft would constitute the first use of a minicomputer facility dedicated to the task of optical navigation performing star detection and satellite center-finding.

The physical configuration, functional configuration, and satellite image extraction algorithm of Linda's Optical Navigation Image Process System are shown below in Figure 2, Figure 3, and Figure 4 taken from the first paper to be presented about the Voyager Encounter Orbit Determination at AAIA Aerospace Sciences Meeting in January of 1980 (Campbell 1980).

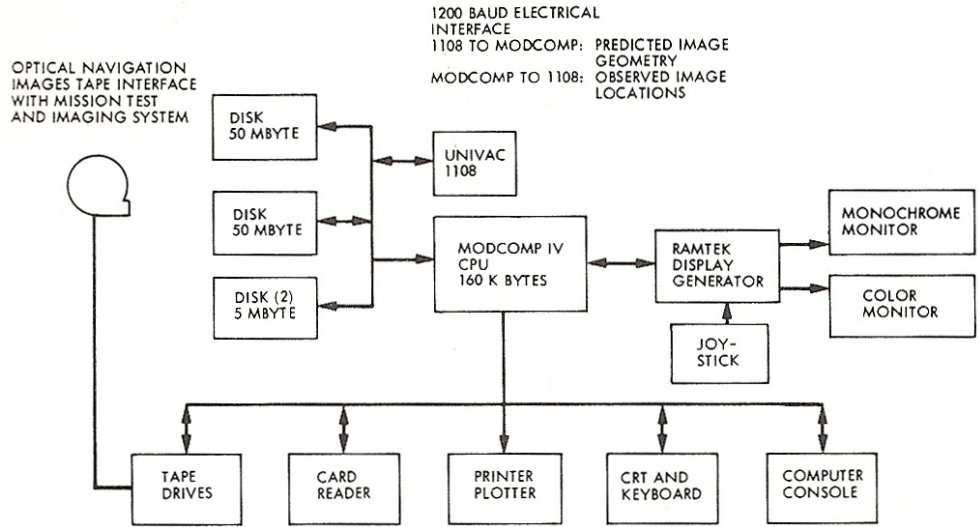

Figure 2. The Optical Navigation Image Processing System (ONIPS) Facility

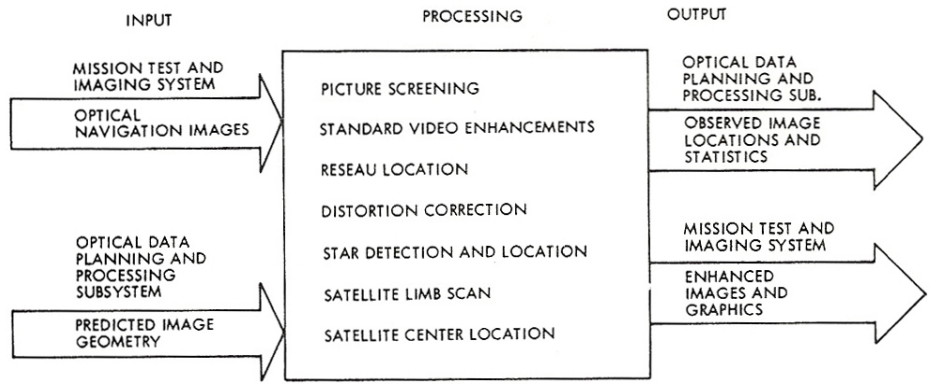

Figure 3. The ONIPS Functional Configuration

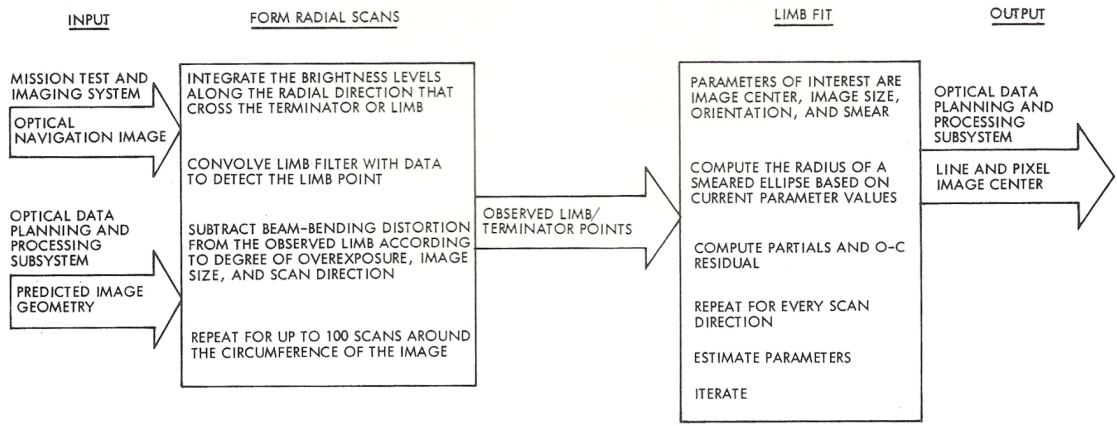

Figure 4. Block Diagram of Satellite Image Extraction Algorithm

## 3. The Process of Scientific Discovery

Reflections of a Young Scientist

Those of us performing the navigation for Voyager were in a team.  We were all members of Frank Jordan's Section at JPL, and one of the seven groups within it.  I was still working under Jim McDanell, but for Mission Operations for Voyager all of the navigation engineers fell under a different hierarchy.

Our Navigation Team Chief was Dr. Edward McKinley.  Ed was a very formal man, and always looked dressed up even in leisurewear; immaculately pressed plaid shirts and kaki pants, which he wore routinely to work.  He had a large smile that was still somehow projected formality.

Second in command was Ed Tavers.  Ed always looked so comfortable that I felt like he was walking across his living room every time I saw him rushing here or there in the Navigation area.  He liked to wear cardigan sweaters and loafer shoes.

The optical data from Voyager, pictures taken by the spacecraft, were not the only data type used in the navigation of Voyager.  This is a spacecraft-based data type.  The pictures were taken by a camera onboard the spacecraft.

The optical data was combined with radiometric data, an earth-based data type, in which radio signals are sent from the Earth to the spacecraft to determine distance to the spacecraft called range and doppler, which is relative motion along the line-of-sight.  Both types of data went into the spacecraft orbit determination process.  Both types of navigators, optical and radio were present on the team.

Jim Campbell and Steve Synnott headed the orbit determination effort; in long hours combining radiomentric and optical data solutions to produce the final spacecraft orbit determination. Jim Campbell was an extremely friendly person, with an infectious smile, whom I admired for his diligence, but did not know well.  Steve Synnott was an avid bicyclist and outdoorsman.  He would later become known for the discovery of many moons of the outer planets in Voyager data.

I respected Steve for his technical expertise.  I was cognizant over the Optical Navigation Image Processing System, a MODCOMP IV minicomputer, which was housed within the Navigation Team bullpen area in the second floor of Building 264 on Lab. Joe Donegan contributed immensely to the creation and maintenance of the ONIPS software. Joe Donegan was an intelligent, thoughtful, and refined man. So, there were only three members of the Navigation Team (Figure 5 and Figure 6) working with the optical navigation data extraction, Joe, who reported to me with respect to ONIPS, and Steve, who was above me.

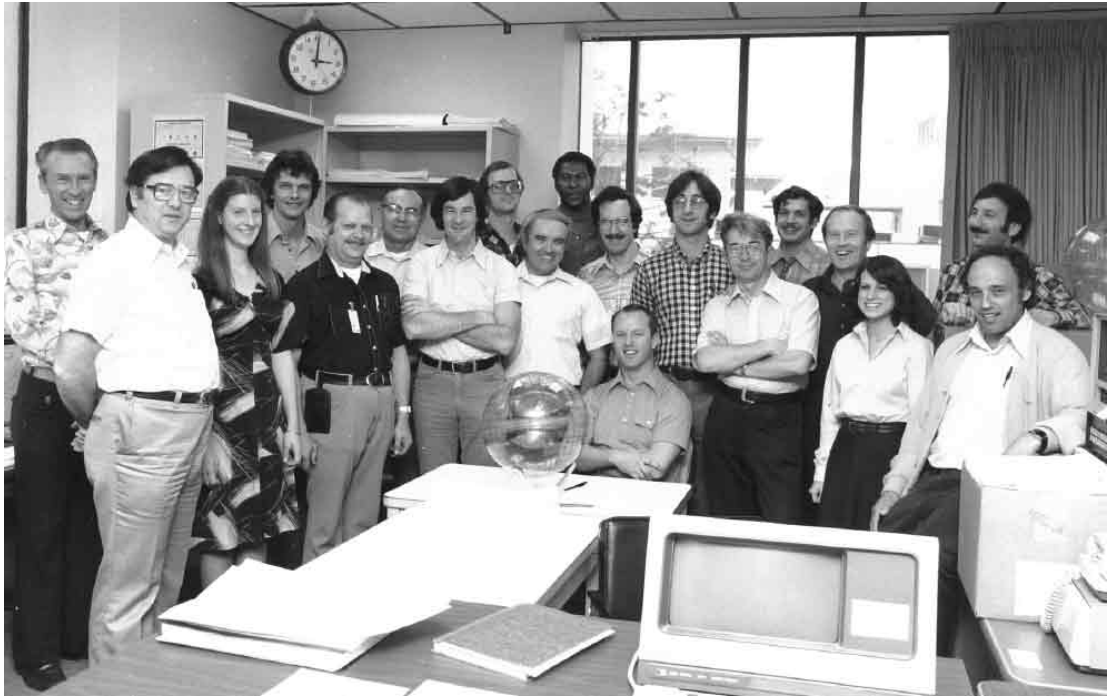

Figure 5. Responsible for Voyager Navigation System Design were: Charles H. Acton, Marvin H. Bantell, Carl S. Christensen, David W. Curkendall, John F. Dixon, Jordan Ellis, Donald L. Gray, Tom W. Hamilton, Claude E. Hildebrand, Robert A. Jacobson, Jeremy B. Jones, Charles E. Kohlhase, Tomas A. Komarek, James P. McDanell, Edward L. McKinley, Lanny J. Miller, Neil A. Mottinger, George W. Null, and V. John Ondrasik, Robert A. Preston, Gary A. Ransford, Gary L. Sievers, Richard H. Stanton, Francis M. Sturms, Jr., Richard E. Van Allen, F. Bryant Winn. Members of the Voyager Navigation Team were: Mark J. Adams, Julian C. Breidenthal, James K. Campbell, Robert J. Cesarone, Leonard Dicken, M. R. Dodds, Arthur J. Donegan, Donald L. Gray, Donald W. Green, Jerry D. Hyder, Robert A. Jacobson, Roger E. Koch, Van W. Lam, So Bing Ma, Stanley Mandell, James P. McDanell, Edward L. McKinley, Margaret M. Medina, Linda A. Morabito, Charles F. Peters, Joseph E. Riedel, George C. Rinker, Lawrence E. Ross, Herbert N. Royden, Andrey B. Sergeyevsky, Stephen P. Synnott, Anthony H. Taylor, Edwin S. Travers, Richard E. Van Allen, Donna M. Wegemer, F. Bryant Winn, Tricia Wood, and Robert R. Wynn, some of whom are pictured here.

Credit NASA/JPL

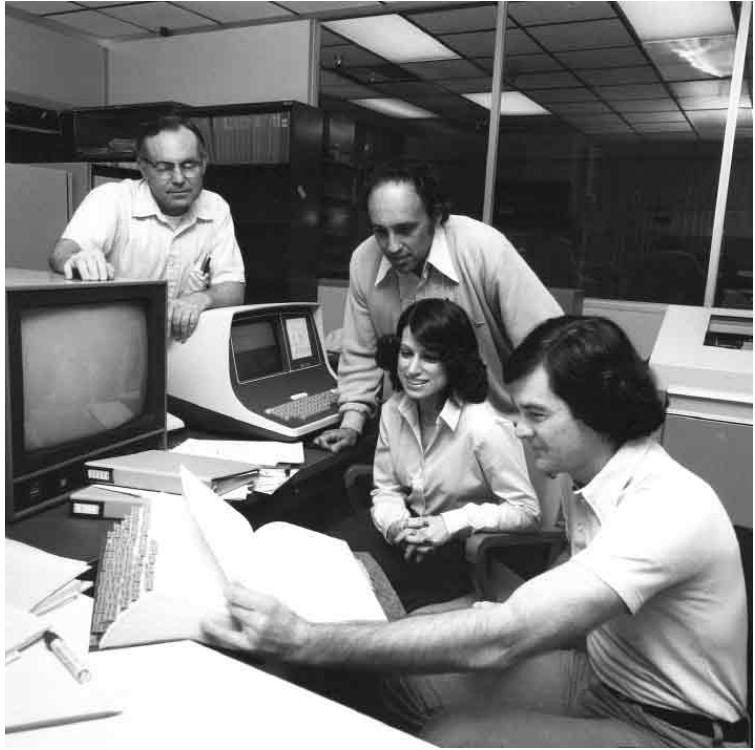

Figure 6. Joe Donegan, Ed Travers, Linda Morabito, and Steve Synnott in the
ONIPS room where the discovery of active volcanism on Io took place.
Credit NASA

There were two areas partitioned off within the rectangular shaped bullpen that held the entire team and their desks. ONIPS was housed in a corner opposite Ed McKinley's office as Navigation Team Chief.

My desk was right outside the door to the ONIPS glassed-off area, and Steve Synnott's desk was right behind mine. The ONIPS area was filled with the central processing unit of the minicomputer, two computer monitors, a magnetic tape reading device, hard-disk storage device, a printer, a modem and other equipment. The minicomputer had to be cooled. Cold air blasted from beneath the floor twenty-four hours a day, in a cacophony of horrendous noise. Over time, dressed warmly, I got used to the wind blowing and the deafening noise within the room.

Above us on the third floor of Building 264 was the Science Imaging Team. This was a team of investigators from around the country, gathered at JPL to carry out scientific studies through pictures taken by Voyager for science. The team was headed by Dr. Brad Smith of the Department of Planetary Science at the University of Arizona. Deputy Team Leader was Dr. Larry Soderblom, a geologist at the U.S. Geological Survey in Flagstaff, Arizona. Larry was very businesslike, but friendly as well.

I came to work to perform the optical navigation image extraction at all hours of the day and night throughout February of 1979, instead of any sleep, and often before the sun would rise. I walked along the path from the entrance of JPL, on a pathway that led to the second floor of Building 264 at the rear.

Just above it, higher up the hill was the SFOF building, or Space Flight Operation Facility Building, the proverbial one that people associate with mission control – each engineer at a named station in front of a computer. The very first light of day was emanating from the East.

I watched my feet, able to see them through the near darkness one such dawn mid-February as I made my way to the Optical Navigation Image Processing System in Building 264. Very suddenly and surprisingly, I recognized something familiar about this moment of being completely alone crossing the Laboratory.

This was the dream that Ray Bradbury who had spoken at USC when I was an undergraduate there had talked about. This was what I had fallen asleep thinking about every night for four years of college and all the years before that. I wasn't myself going into space. But, my life's mission was in space as I walked here.

A small spacecraft representing humanity in the realm of Jupiter was what brought me to this destination, to perform the critical Optical Navigation for the spacecraft at the hours when people normally sleep, when astronauts might dream of a next day's flight. As I looked at the beauty around me, in stars fading from view, at that still, breathtaking moment in time at JPL, I knew my life's path had intercepted my dream.

Getting Things Started

By the spring of 1978, I reported to a new group supervisor Dr. James McDanell at JPL. Something exceptionally noteworthy had happened. After all these years, Chuck Acton, whom I had first met five years before when first arriving at JPL made his decision to pursue another opportunity. There was an opening now for Cognizant Engineer for the navigation of the Voyager spacecrafts using pictures taken by the spacecraft. The Optical Navigation Image Processing system, a dedicated minicomputer system, also known as ONIPS, was losing the person in charge of bringing the project to completion for the work of navigating Voyager.

I was offered the position he vacated of Cognizant Engineer of the Optical Navigation Image Processing System in the summer of 1978.

I accepted the position and began the intense work of getting the software set in shape to perform the countless intricacies associated with finding the centers of the objects in the pictures of the moons of Jupiter against a star background that the Voyager 1 spacecraft would take. Some of these pictures would over-expose the moons of Jupiter in order to be able to see the much dimmer stars in the pictures. This would result in something called geometric blooming or artificial enlarging of the appearance of the size of the moons in each picture.

These pictures of the moons of Jupiter against a star background were translated eventually into the position of the spacecraft with respect to the moons, and were combined with other kinds of data, and then ultimately to TCMs or Trajectory Correction Maneuvers that would set the spacecraft along a corrected and accurate path.

Days ahead would be filled with this type of analytical work based on mathematical modeling of empirical observation in test data, and working closely with Joe Donegan, the ONIPS Cognizant Programmer, to make sure the uncounted equations in this program set were all accurate, as well as getting to know the many pieces of ONIPS hardware that filled an entire room within the Navigation area for Voyager. I began this work eagerly. I had no idea where this work would eventually lead me.

The Heart-Shaped Feature

The discovery of the volcanic activity on Io took place ironically four days after the Voyager 1 encounter with or arrival at Jupiter, after the excitement of the encounter had died down, and most people undoubtedly believed that the wonders that Voyager had showed us in the realm of Jupiter had already been revealed.  This is the historical account of that day, March 9, 1979.

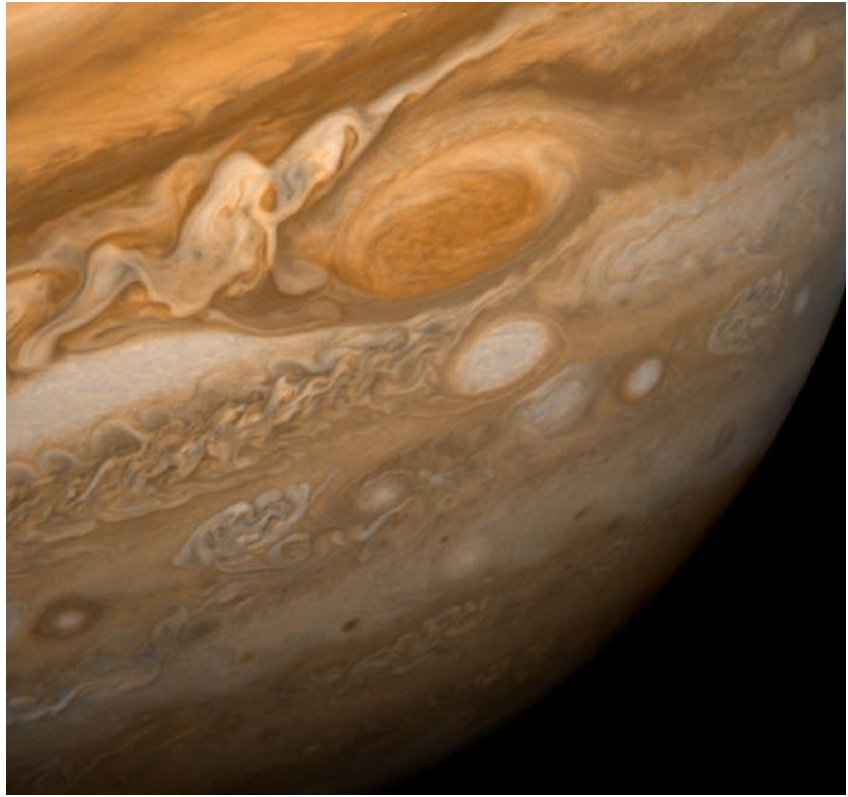

Figure 7. Voyager 1 took this picture of Jupiter on Feb. 25, 1979, when the spacecraft was 9.2 million kilometers from the planet.  Cloud details as small as 160 kilometers across can be seen here.
Credit NASA

Jupiter is a gargantuan world, and the cameras on Voyager began to see more and more of Jupiter as the spacecraft approached the planet in January and February of 1979 as shown in Figure 5.  Jupiter has no solid surface.  Jupiter is a gaseous world that has clouds instead, storms, and lightening.  The biggest storm that has been raging for hundreds of years, the Great Red Spot, would hold the diameters of several Earths within its perimeter.

But, now as Voyager approached we were seeing the cloud patterns at a level of detail beyond what anyone in the history of Earth had ever seen as shown in Figure 7.  These visions of Jupiter would appear on the JPL site-wide monitors, as the pictures came back from the spacecraft.

What human artist could have drawn these cloud patterns?  None could, anywhere at anytime in art history; the intricacies, the swirls, points and folds composed of outer planet material as the Artist's medium of expression; the beauty nearly beyond the mind's capacity to absorb.

I would glance up at a monitor to see Jupiter, rushing in the performance of my work, but stopping because I felt compelled to stop, absorbing what of the beauty of Jupiter's cloud patterns my mind would allow me to absorb, and thinking once again, no human artist…

And then there was Jupiter's moon Io. No one could have imagined Io; no one could have drawn Io either or conceived of how Io would look, with very small exception. No one could have envisioned a real world where its surface might look like an orange kept too long and covered with a burgeoning, whitish parasitic mold, or a bizarre pizza pie where the chef got carried away with too many varieties of cheese, some tinged in blue as seen in Figure 8.

I'd never seen this world before, the way its surface looked. No one had until Voyager encountered Jupiter and showed us. Galileo had seen Io in 1610, but as a dot in his telescope on Earth circling another planet, with three other dots doing the same thing. These are the Galilean satellites of Jupiter: Io, Europa, Ganymede, and Callisto, discovered by Galileo.

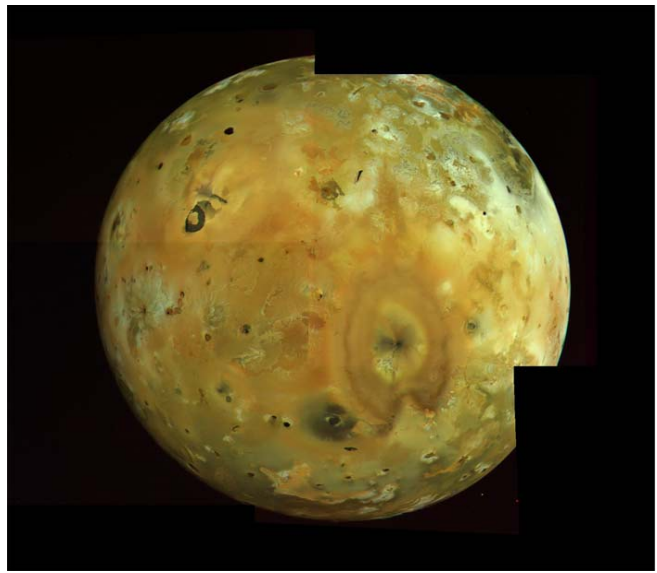

Figure 8. (Original Caption Released with Image) Perhaps the most spectacular of all the Voyager photos of Io is this mosaic obtained by Voyager 1 on March 5 at a range of 400,000 kilometers. A great variety of color and albedo is seen on the surface, now thought to be the result of surface deposits of various forms of sulfur and sulfur dioxide. The two great volcanoes Pele and Loki (upper left) are prominent. Image/Caption Credit: NASA/JPL

These are planet-sized worlds that held a secret even beyond the one that confined Galileo to house arrest for the rest of his life after seeing them. Not only did they prove in 1610 that the Earth was not the center of the universe, since they orbited around another center, violating religious doctrine at the time, but the existence of two in orbital resonance with Io made Io into something beyond imagination.

On very few occasions did I return home for any amount of rest, no more than four hours a night during the encounter phase of Voyager 1 with Jupiter in February of 1979. This was true right up to the encounter with Jupiter on March 5, 1979.

On March 4, at 11:20 AM, I walked down the stairs of my condominium in Pasadena at home preparing to return to work, and caught a glimpse of a press conference emanating from JPL, where I would be within the hour.

My images of Io for Optical Navigation were over-exposed to capture the delicate and much fainter star images in the frames. I could not see detail on the surface of Io in my pictures and was working too hard for navigation at that point to concentrate on what the scientists were seeing in their color pictures of Io.

I caught the image of Io on my television as I came down the stairs at home. I stopped where I was. Amid the reddish and whitish hues of this inconceivable "moldy" looking world, was a massive heart-shaped feature.

Scientists on the science imaging team were proposing that the surface of Io might not be too old in geological terms, much younger than the four billion years of the moon's existence, and much younger than the surface of other moons, because this surface was smooth and not cratered. I could feel tears begin to roll down my face at the sight of a world more unexpected than imagination itself, as I looked upon that ugly heart-shaped feature. I would remember this later. It was as if I had been gazing into destiny.

The Discovery of the Volcanic Activity on Io

On the morning of March 9, 1979 when I woke up, I had actually considered not coming into work. This was four days after the completely successful Voyager 1 encounter, or arrival at Jupiter. I was very far behind on my rest. During the 28 days of February, the data had been pouring down on us like rainfall. There would be times when I knew I had exactly four hours out of the next twenty-four to sleep, and I simply couldn't sleep at all because I was just too nervous and excited about the importance of the work I would be doing for the navigation of Voyager.

I knew the frames I would be processing today would be used for something called post-encounter satellite ephemeris reconstruction. The spacecraft was looking back over its shoulder for one more look at the satellites of Jupiter, and we would use this data to better improve our knowledge of the orbits of the satellites. There was no timeliness constraint on this work, and it was not for the post encounter Trajectory Correction Maneuver, consequently it was not extremely important.

Still my dedication to the work was so deeply ingrained that the thought of not coming in was just wishful thinking, a sort of daydream about more sleep after more than a month without it.

I arrived at the Nav Team area at my desk at 9:15 AM, March 9, 1979. The MTIS QEDR tape containing the four pictures I would be processing today, the first of these post-encounter frames, was already on my desk. It had probably been placed there by a runner at 6:00 AM. The pictures had been taken by Voyager 1 the day before.

I was aware in advance there was a high phase angle in these pictures. What that meant was the angle between the direction or line-of-site to the spacecraft from the moon, viewing direction, and the direction or line-of-sight from the moon to the Sun, incident sunlight, which would light the moon, was very high. This was the viewing geometry. Moons and planets do not shine by their own light. They shine only by reflected light from the Sun, unlike stars. Our Sun is, of course, a star.

When we see a full moon in the night sky from Earth, it is because the Sun is directly behind us as we are viewing the moon, and directly opposite the moon. That full moon on Earth represents a very small or zero phase angle; the angle between the viewing direction of the moon, and incident sunlight.

The smaller the phase angle in any viewing geometry, the larger portion of the moon that is lit; hence a zero phase angle leads to a fully lit, or full moon. With such a high phase angle, approximately 124 degrees in the viewing geometry of these post-encounter frames, the moon of Jupiter would appear to the spacecraft camera to be far from fully lit, in fact, like a crescent or crescent moon.

With only a partial limb or edge of the moon lit, unlike the full moons we had used for the orbit determination, the computerized limb-fitting process would be undermined and impact the accuracy of my center-finding and therefore the usefulness of the data.

By the time I got organized and copied the pictures to ONIPS' 50 megabyte disk pack, it was 10:00 AM. I displayed the first picture on ONIPS' monitor. There was nothing in it, it was devoid of a satellite. This had never happened before in an Optical Navigation picture, to have missed capturing one of the images of a moon of Jupiter. I was very surprised, and I quickly considered what might have happened would be a pointing problem, the direction that the camera was looking at the time was inaccurate, or a timing problem on the spacecraft. If the time the picture was taken was incorrect, then the geometry was changing so rapidly with the spacecraft moving out of the realm of Jupiter, the moon could easily have been missed.

As I displayed the second picture, Steve Synnott walked into the ONIPS room. The picture was of Io. Io was in fact a crescent, just as it would be with a known high phase angle as we were expecting in these post-encounter frames.

The dark side of Io, the unlit side of the crescent, the larger portion of the moon, was also visible though just faintly above the blackness of space in the picture. This is because light from the Sun was bouncing off Jupiter and back onto the dark side of Io's surface. This happens with Earth's own moon as well. When Earth's moon is a crescent because of the geometry of the viewing, light will bounce off the Earth and back onto the moon's dark portion not illuminated directly by the Sun. This dark portion of our moon can therefore be seen above the blackness of space, but fainter than the crescent itself.

Steve and I looked carefully at the crescent image of Io. We both knew that our center-finding algorithms depended upon a fully lit satellite to do high-accuracy work. There was no favorable geometry for viewing the satellites of Jupiter as the spacecraft looked back over its shoulder one last time, as it flew by. Yet, these frames were all that was available to us dictated by the circumstances pulling away from Jupiter.

Steve was always highly focused on whatever he was considering in front of him. I watched his eyes zero in on the Io frame displayed on my monitor. Neither of us at this moment was showing outward signs of exhaustion, but had either of us had any demanding work of any kind, we both knew that exhaustion was within us both. The relative unimportance of this work matched a kind of automatic and after-the-fact following through state we were both existing in after the pre-encounter work that had the intensity of a hurricane.

> That thin crescent of Io reaffirmed as we gazed at it that all of these post-encounter pictures were essentially useless. Steve and I agreed on it. Steve got up and left the room.

Regardless of its usefulness, I got to work on processing the picture of Io because I didn't want to leave a job undone. I performed what is called the registration of the picture, in which I aligned a graphic overlay of the way we had planned this picture, what we expected to see in it, interactively using a joystick. Joysticks are the way children control their interactive game systems. They move virtual spacecrafts or cars around video games with a hand-held controller, which can occupy kids for hours. Joysticks were not that commonplace in 1979, and my computer system was hardly for gaming.

Such a graphic overlay always allowed me to know where to start looking for the star images in the picture, which were too faint to be seen without some computer enhancement to bring them out through image processing. The two stars predicted to be in the picture with Io were a 7.2 visual magnitude G5 star, which refers to its brightness and spectral class, which is a temperature and color indicator, and a fainter 8.3 visual magnitude G5 star. The star names or identifying numbers represented a catalog designation of AGK3-2006 and AGK3-10021 respectively, although the usefulness of that information is basically just to designate them with a name.

I moved the interactive graphic representation of Io, which was just a computer-generated circle of the correct size over to the actual crescent of the moon on the monitor using the joystick control.

Given that, I knew the 7.2 visual magnitude star fell into the upper left hand quadrant of the picture. So, I needed to zero in on that quadrant to do some image processing and bring out the faint image of the star. It is the actual location of the star and satellites in these pictures that yield the information I was looking for.

Before the fact, or what is called apriori, I could only approximate what the pictures would look like, based on a location of the spacecraft and an orbit of the moons that all needed refining. The true positions of the moons and the stars in the Optical Navigation pictures held the information for those results to be eventually derived.

In zeroing in on that upper left hand quadrant of the picture and with a linear stretch to enhance the contrast in that quadrant, the image of the star jumped out at me, and became brighter than the darkness of space that surrounded it. Yet, it was very far from the computer generated graphic of a little box that had this star's name on it, on the overlay.

Since I had placed the overlay directly upon Io, I was once again very surprised by one of these post-encounter frames to see that the star appeared so far from its predicted or apriori location. In all Optical Navigation frames I had processed for Voyager 1, the positions of the stars were usually better predicted by the navigation programs. Once again, I thought that the picture might have been taken at a different time than planned, or some kind of an orientation problem existed in the direction the camera was pointed, or perhaps the overlay itself had been drawn with an older spacecraft predicted position.

Even though these frames and this work was low priority, I did not feel immediately comfortable with any of these explanations as to why the star would be so far from its predicted location, and given the fact we had missed the satellite in the first picture completely, everything about these post-encounter frames seemed foreign to me, nothing like the precise and flawless work that I had done pre-encounter which had been part of the spectacular success of Voyager 1 at Jupiter.

I was thinking I was relieved on the one hand these frames were unimportant, but upset on the other hand that my desire to do this job was being undermined by these various peculiarities. I thought, "Oh well, I'll go and look for the other star."

It was at the moment that I displayed the lower left hand quadrant in which the 8.3 visual magnitude star was predicted to be in and into which the image of Io mostly fell, and I performed a linear stretch to enhance the dim signal of the star, that I noticed the peculiar marks to the left of Io. I have never since to this day remembered anything about the star I was looking for after I performed the enhancement that brought out the bright marks beside Io shown in Figure 9.

I succinctly asked myself, "What's that?"
In the same way, not aloud I answered myself, "It looks like a satellite behind a satellite."
It did indeed appear, this large bright anomaly like another moon peaking out from behind Io.
My mental process wouldn't stop, and I was insisting to myself in the absence of any new information that I find out, when I asked myself immediately again, "What's that?"
Once again, my mind automatically concluded, "It looks like a satellite behind a satellite."

Since I couldn't know at that instant, what was clear was that we had captured something unintended in this picture and that is always exciting! I jumped up and went to get Steve. Steve came very quickly to the monitor. When he saw it, this good-sized anomalous protrusion beside Io, Steve said:

"Jesus, what's that?"
We both agreed it looked like another satellite behind Io.

Steve went to get one of his data products, a Trajectory Geometry Predicts output. Characteristically, he pushed the pages of the oversized printout back one and then the next until he found where he wanted to be looking. At the time this picture was taken, there were no other of the Galilean satellites predicted to be anywhere near Io, not even the little moon Amalthea, which would be only 5 pixels in diameter at the distance of the spacecraft at this time. This object, if indeed represented the full sphere of a moon, would be approximately 20 pixels in diameter.

I told Steve that the timing of the picture might be wrong, what I had been thinking prior to finding the anomaly. I had noticed the time tag or what was called the FDS count of the picture was one count off, predicted from actual. The predicted was 16481.10, and the actual 16481.9. I had heard that the spacecraft was having a timing problem after the encounter. Perhaps this was an indication of that, I went on.

Then, I thought perhaps the large crescent satellite in the picture wasn't even Io; as in all my pictures, most of the crescent was saturated, or overexposed, and no surface detail was visible. But, then I realized this moon crescent was indeed the right size for the graphic overlay I used in the registration process, so that meant this was Io and not one of the other Galilean satellites.

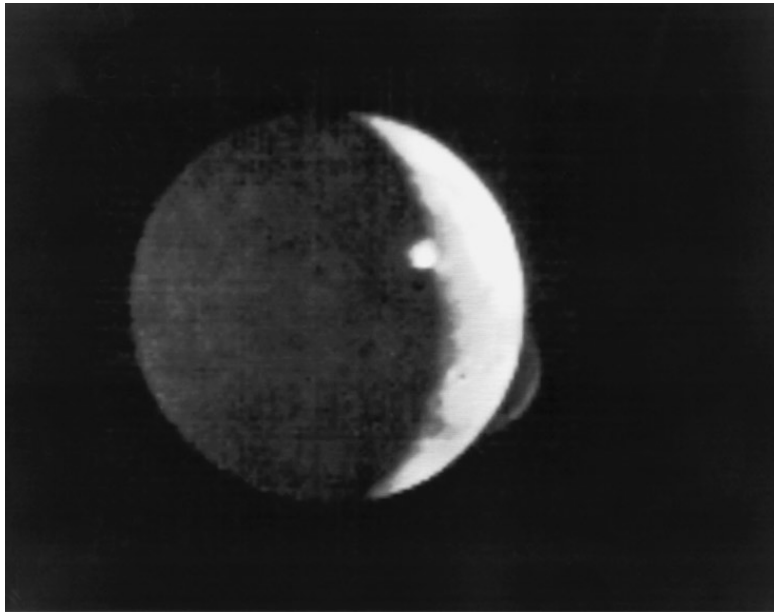

Figure 9. (Original Caption Released with Picture)This dramatic view of Jupiter's satellite Io shows two simultaneously occurring volcanic eruptions. One can be seen on the limb, (at lower right) in which ash clouds are rising more than 150 miles (260 kilometers) above the satellite's surface. The second can be seen on the terminator (shadow between day and night) where the volcanic cloud is catching the rays of the rising sun. The dark hemisphere of Io is made visible by light reflected from Jupiter. Seen in Io's night sky, Jupiter looms almost 40 times larger and 200 times brighter than our own full Moon. This photo was taken by Voyager 1 on March 8, 1979, looking back 2.6 million miles (4.5 million kilometers) at Io, three days after its historic encounter. This is the same image in which Linda A. Morabito, a JPL engineer, discovered the first extraterrestrial volcanic eruption (the bright curved volcanic cloud on the limb). Jet Propulsion Laboratory manages and controls the Voyager project for NASA's Office of Space Science. Image/Caption Credit: NASA/JPL

As I expressed these thoughts to Steve, I became suddenly aware of the level of exhaustion I felt. What was happening here was no longer some form of routine following through with rudimentary work. Steve was wearing his everyday striped polo shirt and casual pants. I had on one of my everyday shirts and skirts that kept me warm inside the ONIPS area. But, each of us was slowed down, belabored by the simplest acts of moving around. Steve looked spent. I knew I was too. I realized almost horrified that I was having trouble thinking. Steve and I had talked about this yesterday. He said his brain felt numb.

Despite that, Steve managed a quick back-of-the-envelope computation. An object at this range or distance from the spacecraft, which was four-and-a-half million kilometers away, of this apparent size, if it indeed represented part of the full sphere of a world, would be a world with a diameter of one thousand kilometers. Such an object would not have been able to hide from discovery from the Earth. It would have been seen and discovered by an astronomer somewhere on Earth, as Galileo had discovered Io. Steve caught his breath. He said, "It can't be a satellite. I would be ridiculous and impossible not to have been seen from Earth."

I responded:

"Then it's an Io flare."

I was joking of course. A disturbance on our star, the Sun, produces flares that rise very high off the surface of the Sun. Since Steve had eliminated the possibility of the anomaly being a newly discovered moon of Jupiter, then my mind had done the logical thing and correlated the phenomenon whatever it was with Io itself. Yet, the protrusion that otherwise did look spherical, was gigantic with respect to the size of Io, about fifteen percent the radius of Io. It was too large a phenomenon to be emanating from Io.

Psychologically, your brain can't look at a saturated, overexposed crescent, an entire world like the Earth in crescent lighting and decide that a hefty good-sized object off its limb or edge has anything at all to do with that world. It's too big, too mind-boggling a possibility. Not even a tired brain could conclude that by looking at the anomaly I had found.

Steve listened. He said, perhaps pictures for science planned by the Science Imaging Team were also taken at that time, and the scientists above us on the third floor were on the verge of making some kind of a discovery as well. He said he would look in one of the sequencing products to see if any other imaging had gone on at this time. Steve left the ONIPS room.

What Had Not Been Seen Before

I was now alone with the image of Io and the anomaly. I sensed the significance of what was in this room. I knew I had something here. I knew it was important. I sensed that I was seeing something that no human being had ever seen before.

My heart pounded.
For the first and honestly only time since, I allowed myself some genuine heart-pounding excitement.
In the long term, the significance of the discovery would come to far overshadow anything else.
In the short term, I had a feeling about how difficult the hours would be ahead.
This was my time with my discovery; the only time that has ever belonged to me and this discovery.
And, I cherish it. It stays with me always.
It was the stuff of dreams.

I paced back and forth. I glanced at Steve through the windows that encased the ONIPS area to watch his progress as he spoke to the Navigation Team Chief, Ed McKinley, and appeared to be working. However, I saw Steve leave the Navigation area.

Fifteen minutes passed since he had seen the anomaly in the picture of Io by the time he returned. The first thing I noticed was that he was carrying a carton of milk in his hand. I was thinking, carton of milk… The only place I knew anyone could get a carton of milk this time of morning was the JPL cafeteria. There seemed to be a disconnect between what I was thinking about this image of Io and taking a trip to the cafeteria.

I rushed out of the ONIPS room over to Steve.  What had he checked?  What had he found out? I questioned.  Had any other imaging for science gone on at that time this picture was taken?

Steve said, quite deliberately, "I haven't been working on it.  I haven't talked to anybody.  Look, I'm not very excited about this.  It could hardly be an undiscovered satellite with a radius permitting observation from Earth."  He went on to say therefore he had dismissed the image entirely after having left only minutes before.

My reaction to this was personal shock.  Steve was all about technical expertise everyday I had ever known him. I had never before considered the possibility that any of the people who surrounded me with their profound technical backgrounds like Steve, could be any different than I was in the sense of their grasp of the rarity of an anomalous observation.

I reminded Steve he was about to determine if any other imaging had gone on at the time of this picture.  Steve seemed willing to do that.  Steve located a final sequencing product.  We determined that no other imaging had gone on at that time.

We went on to explore the timing problem that I had mentioned about the picture in trying to explain the various things about it that seemed unusual.  As we talked about it, I recalled this had happened on one if not two other Optical Navigation frames last month, to have an FDS count differ from predicted, and in the heat of the work and the time constraints I had neglected to mention it to Steve.  That seemed crucial now in the context of an oversight that might have caused an error, albeit, a small one in our navigation solutions.

It was important that Steve and I consider this immediately.  It turned out the error had affected only one previous Optical Navigation picture, and it was important to correct our future solutions for this one picture.

When our attention returned to Io, we discovered that the frame had indeed been shuttered 38 seconds earlier than predicted, 1979 March 8, 13 hours 28 minutes Greenwich Mean Time.  This could still in no way account for the presence of the anomaly, although it completely explained the larger than usual difference between the graphic overlay and the relative positions of Io and the one star I had seen.

Even though we were working on the image, Steve and I seemed worlds apart in our impression of this image.   I decided to do something on my own.

Narrowing Interpretations

I telephoned Tom Duxbury. I wanted to make sure that what was appearing beside Io was not merely an artifact of the camera.  There were such things that appeared in every picture taken by spacecraft vidicon cameras like Voyagers', if you knew where to look for them.  The Viking "cheerio" as we had called it during the Viking mission was a prominent artifact in one of the Viking orbiter cameras and was about the size of this object besides Io.

The object beside Io, the anomaly, itself had the properties of a crescent, more brightly lit around the edges; a little like a ringed cheerio of breakfast cereal notoriety, which is how the Viking cheerio had appeared.  Tom confirmed on the phone what I had noticed during the encounter, that there was in fact a Voyager cheerio too, but that he thought it was closer to the upper left hand side of the field of view.  This reaffirmed what I remembered as well.

I told Tom I was seeing something in an OPNAV frame I didn't understand.

Tom said, jokingly:

"Perhaps you have discovered a galaxy with a hole in it."

I was relieved by his humor, and the conversation ended that way. It dawned on me now, that I had no control over this situation whatsoever, and how rapidly my ability to continue to investigate the anomaly could be ended.

It was the larger issue that I knew of no path to walk on to have this situation proceed as it should. I had no prior knowledge of a discovery coming out of an engineering setting at Jet Propulsion Laboratory. The Science Imaging Team made science discoveries. What path could recognize ONIPS as a viable area from which to push forward to completion an analysis of the image? A path that would result in the preservation of a record of these events, so that what had already happened here, or what was about to happen here; that one day it would be known what had transpired. This path represented the way to survive this experience for me, and I knew of no precedent for it.

Two hours had passed now since I had first seen the anomaly. However awkwardly, Steve and I had now explored all of the timing and technical geometry considerations of the frame. We were still sparring about Steve's seemingly indifference to my quest to investigate the image further, when the subject of what to do next came up.

He said, "You have done everything you can," and he said to call the Science Imaging Team.

I placed a call to one of the science imaging team members who had utilized my help to see the ring of Jupiter for the first time clearly on my computer system only 5 days before. The administrative assistant in the Science Imaging Team area was a friend of mine and she told me she would leave the message for the particular team member to call me. She told me as well that he might be at the Image Processing Laboratory across the street.

I placed four calls to various numbers at that facility and each person who answered there had never heard of the science team member I had been told to reach. They all said, "Who?" rather indignantly.

I looked around when I'd gotten off the phone. Steve was gone. I presumed he went back to the cafeteria. I was right. It was after twelve and lunchtime. I had not eaten all day and had not eaten much the day before. I was fighting back a good measure of physical weakness by this point, but nothing was going to stop me now. I had to do something.

I placed a call to Peter Kupferman, an astronomer, who interfaced a lot with the Imaging Team, but was not a team member. I did not reach Peter with the first call. He called me back minutes later. I told Peter I was seeing something strange in an OPNAV frame.

Peter, if anyone, could conclusively eliminate the artifact theory, as he was a camera expert. He was the one person I knew who could determine if any quality of the camera could induce the appearance of the anomaly, thus not representing something Voyager was actually imaging. Peter told me he would come right down, and he showed up a minute later. I pointed to the image on my monitor.

Peter said, "Oh my God!" and moved very close to the screen.

He asked me if he could call Andy Collins, the Voyager Experiment Representative.

Andy Collins was known to us Optical Navigators as a good guy, because he was. His job was to interface with the scientists as the bridge between the engineering world at JPL and their team. Andy realized if there is no Optical Navigation there is no science! I told Peter it was okay to call Andy.

Andy Collins, who had a wonderful boyish quality, came down to the Navigation area eating a cheeseburger. I do not believe that "Oh my God!" was a quote for Andy, but that is what his face was saying. It was the most incredible experience watching Andy Collins look at the image of Io. He was acrobatic about it.

Andy sat on top of a table in front of the monitor, crossed his legs, looked at the monitor and at one point he lay down on the table on his belly and looked at the monitor, all the while taking bites out of that cheeseburger; his face still spelling out, "Oh my God."

I watched Andy and the cheeseburger. I have never been so jealous in my life of anybody having anything than I was of Andy having that cheeseburger. I envisioned headlines that might be written, as I watched this amazing sight, that perhaps I would perish from hunger while pushing forward to find out what this anomaly in the picture of Io was.

What Andy did say verbally first, was that he would need digifax hard copy of the picture. This told me that the picture would leave the ONIPS area. I protested that I did not want the picture to leave the room, because honestly I had not had the opportunity to take my investigation of this image as far as I knew I could. He seemed to understand. But, Andy said that ONIPS was not capable of the type of image processing that needed to be done; that the Image Processing Laboratory facility across the street could do. I told Andy my system could do anything that IPL could do, which was of course not true and I could not appreciate suggestions of taking this out of my hands.

We all sat down, Andy Collins, Peter Kupferman, and I and I performed some basic "look-see" capabilities that ONIPS could do on the image. Peter definitively told us that what we were seeing was real, that whatever this anomaly was, was not camera induced.

The largest single revelation that came out of this work was from investigating the portion of Io's surface that was receiving reflected light off of Jupiter, the portion darker than the bright crescent of Io. The brightness or DN level (Density Numbers ranged from 0 for black to 254 for white associated with each pixel in the frame) of the dark portion of Io, was at the same level or brightness as the dark portion of the anomaly; the portion of the anomaly that was also not bright, bordered by a brighter crescent shape of its own, as in Figure 10.

This matching of the brightness levels gave credence to this observation as though whatever was happening in space in the realm of Io was in some way receiving the same degree of lighting from the Sun as was Io. Between this and Peter's assessment, this observation was gaining credence.

To deal with something that has never been seen before is to have nothing to equate it to.
First for the human mind, to know it's real.

Figure 10. Original Optical Navigation Image Processing
System data product of the Io Plume Pele in the Discovery
Picture generated by Linda Morabito on 9 March, 1979.
Contour lines by DN range drawn in by Steve Synnott.
*Courtesy Linda Morabito*

Andy suggested taking the picture away again, and I refused again. The rest of the conversation was Andy and Peter taking turns with one another in different expressions of "Oh my God!"

Steve Synnott returned in the midst of all of this. My confidence in his ability returned automatically, from long habit. But, I said, "These people are a little more excited than you were." I looked him squarely in the eye, and asked him in that way to take over. I was headed to get a cheeseburger.

The cheeseburger I grabbed at the cafeteria just past mid-day on March 9, 1979 didn't last long. When I returned to the ONIPS area within minutes of having left to get the only food I had that day, the excitement was spreading.

Andy Collins, Peter Kupferman, and Steve Synnott were wandering in and out of the ONIPS room, along with maybe ten other members of the Navigation Team. There seemed no way to prevent that. I still knew of no path to walk on to know how to survive the challenges that were coming. But, I did sense I was in exactly the right frame of mind.

I was so dug in, that everything that was happening around me seemed peripheral to the sole purpose of me taking this observation to the ends of the Earth to understand it. If that is what it would take, that is what I would do. Tom Duxbury showed up now too based on my earlier call to him. An animated discussion began between those principally involved.

Andy Collins had it divided into three possibilities. 1) A newly discovered satellite of Jupiter; 2) a newly discovered satellite of Io; or 3) an Io-based phenomenon.

Tom Duxbury assured us that an Io orbiter was dynamically possible, in other words, Io theoretically could have a moon orbiting around it even if it were a moon itself. But, Steve

Synnott and I had established so much earlier in the day that we did not feel we were dealing with a satellite.

Tom Duxbury mentioned that the Voyager cheerio was a dark feature and not a bright one, as the anomaly was, and I agreed. So, this was not the most well known artifact of the camera system.

Andy Collins said if it is an Io-based phenomenon we have better find the location of that phenomenon – could it be a cloud? Steve Synnott said coincidentally he had been at lunch with a group of people discussing Stanton Peale's article, Stanton Peale of the University of California at Santa Barbara, which had predicted a volcanically active Io, and had been published before the encounter.

Peter Kupferman said, "If it is a gas shell, this is what it would look like." When Peter said that I recalled my very first impression of the anomaly, or gas shell, if it was, which had its own crescent or way in which the sunlight was hitting it, was more consistent with an Io-based phenomenon, than a body beyond Io, which would in all likelihood be in the same orbital plane as Io. If it were another body beyond Io, the crescents of that body and Io would probably have been more similar in their orientation. That had been my instantaneous impression.

Andy's suggestion was to compute the sub-spacecraft latitude and longitude, which would be at the direct center on the image of the moon, and from there, compute the latitude and longitude of the anomaly on the surface, to see what we would learn. The whole Navigation Team was milling around now.

No one understood what they were seeing, reinforcing the degree of difficulty associated with interpreting this image. Yet they could feel that we understood more although we would not tell them. Some members of the Navigation Team came close to wanting to congratulate me, but things seemed too uncertain.

I caught a glimpse of Steve Synnott talking with the Navigation Team Chief Ed McKinley and Steve's own counterpart on the team, Jim Campbell outside the ONIPS room. Steve appeared to be expressing great discontent and looking at me. I went right over to him and he criticized me for having contacted so many people while he was away at lunch. I informed him I hadn't done that. I had only called that particular member of the Science Imaging Team who had not even shown up yet, Peter Kupferman and Tom Duxbury much earlier. Steve looked completely shocked. He asked me why I had called the Science Imaging Team. I felt shock ratchet through my entire body. I said, "That is what you told me to do!" Steve said, "I did not tell you to do that."

Hearing this, I believed this was some kind of a joke Steve was playing on me, which was the only plausible explanation, and I paused, waiting to hear exactly what kind of a joke this was. Steve said nothing, I couldn't believe it; he could actually not remember telling me to place the call.

I regained some composure, but in an instant ran out of the Navigation area and into the hallway. I slammed the button on the elevator and waited for an elevator car to arrive. I got in it and hit the third floor button. When the elevator doors opened, I walked into the Science Imaging Team area as though I belonged there. My friend was sitting at her desk, where she had taken the message much earlier that day. She said, "Hi, Linda!"

I don't know quite how, but out of 50 mailboxes, my eyes settled on a yellow note slip. I grabbed it like I knew what I was doing. It said, "Call Linda Morabito at extension 5971." I took it with me and no one saw me do it.  I came back downstairs.  Steve knew what it was when I threw it into the trashcan behind my desk, between our two desks.  We sat there in complete silence at our desks.  I was staring into nothing and could see Steve in my peripheral vision doing the same thing for a very long time.

More Come to See

It had been many hours since first seeing the anomaly in the picture of Io.  My analysis of this image hadn't been stopped yet by anything, and although I did not know the path to walk on for survival that day of March 9, 1979, I was still walking, and therefore was being given the chance to find out.

With crowds still milling in and out of the ONIPS room, and Steve Synnott and I sitting there at our desks, Andy Collins walked over and asked Steve if he could call in the Voyager Project Scientist, Dr. Edward Stone.  Steve granted that permission.  Dr. Stone would become the Director of Jet Propulsion Laboratory in 1982.

For Ed Stone's exposure to the discovery that day, and again the next, and every day since, I have nothing but the deepest admiration for him and the most profound respect. Years later, I would watch Ed Stone literally run after a researcher who was compiling information on Voyager and say, "Come over here!  Talk to Linda!"  I didn't fit into the theme of his book and the researcher sort of ignored Ed.  I was delighted, laughing with Lou Friedman about what Ed was doing running after the man, and felt privileged enough to just be looking on.  Although the interactions I have had with Ed Stone have been all too brief over the years, each one affirms the integrity of this remarkable man.

I believe Ed Stone knew what he was seeing in the image of Io right away when he arrived at the ONIPS area, although he did not say that.  His eyes were literally twinkling, with no exaggeration, and he kept repeating to himself, "This has been an incredible mission."

When Ed Stone left the ONIPS area, saying and doing nothing more, the crowd finally dispersed.

Steve Synnott and I got to work.  We were aware of one set of latitude and longitude parameters, for the heart-shaped feature on Io.  It was longitude 250 degrees, latitude –30 degrees.

Steve made a phone call to get the sub-spacecraft latitude and longitude on the Io picture.  It was 338 degrees longitude and –4 degrees latitude.  Knowing the orientation of the picture, and that ONIPS inverts pictures from the spacecraft, I was able to do the computations in my head for an approximate latitude and longitude of the anomaly if it were on a gas shell on Io.  I estimated 90 degrees further east in longitude and 30 degrees further south in latitude.  That put the anomaly at 248 degrees longitude and –34 degrees latitude, too close to the longitude and latitude of the heart-shaped feature to ignore.

This moment evoked my knowledge of history and the story of Jocelyn Bell, who had not been credited for her discovery of pulsars in 1967 at The University of Cambridge in England.  There was no joy in this discovery at this point.  I told Steve of the very real possibility that credit for this find could easily be taken away from us.  I didn't want to love astronomy any less.  I didn't want any more grief right now. Another thing I realized at this moment was that I had looked upon that ugly heart-shaped feature on Io courtesy of our backbreaking navigation only five days before, and had cried. This is when I realized it felt like I had been gazing into destiny.

Armed with this new revelation of a latitude and longitude of the gas shell, we were now hypothesizing it might be, Steve and I set out for the Imaging Team area to find Andy Collins and Peter Kupferman.

It was a fight to get people out of Andy's office so that we could shut the door, but once they started leaving, it is amazing how fast it cleared out. Not many of these people were scientists themselves, fortunately, or they would have wondered about our presence there.

Present were Andy, Steve, Peter, and I. I do not recall Andy's specific reaction to our revealing that the cloud was hanging basically over the largest surface feature on Io, but all of us were gravely serious throughout the meeting.

We established a plan behind those closed doors. Andy had an ingenious idea. The very software that had been used to find the centers of satellite images to a high accuracy on ONIPS, could now be used to limb-fit a volcanic plume, to find its center, thereby determining to a high accuracy the latitude and longitude of the ejection center. I would limb-fit a circle to the edge of the crescent of the gas cloud, and Peter would help me to determine the satellite based latitude and longitude of the ejection center of the cloud as in Figure 11.

Ironically at lunch, Steve had been sitting with people who were discussing Stanton Peale's publication that had predicted an "active" Io. The mechanism of the Io volcanism, as discussed by Stanton Peale in his article, was a gravitational pumping of the satellite induced by the other Galileans.

Andy said ejection velocities and particle sizes of the eruption would have to be determined.

Andy strongly suggested again that he be allowed to make digifax hard-copies of the picture. This time I consented.

When Steve and I returned to the Navigation area, I went into the ONIPS room and I got to work immediately. It was far into the day now. Steve came into the ONIPS room and he was acting very restless, very much an outward manifestation of what I had been dealing with all day.

I probably could not conceal the hope in my eyes that he was there for me on this. He said, "You will never forgive me if I go home." I thought about it and hesitantly replied that I guessed I would understand.

"No you won't," he said, and left anyway. He was right, I didn't.

Testing the Idea

Andy Collins had suggested upstairs on the third floor of Building 264 that Peter Kupferman work with me on the determination of the latitude and longitude of the ejection center of the place the volcanic plume was emanating from. When Peter came back to the ONIPS area, we worked for the next two hours.

As we began, navigation people came in occasionally to inquire. This was a strain on me, because I did not know what would transpire. I didn't know what to expect at any turn.

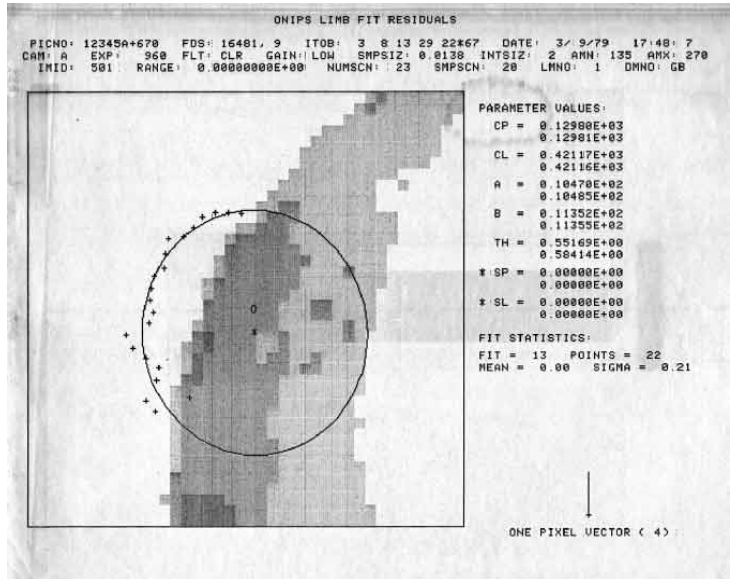

Figure 11. Original Optical Navigation Image Processing System
data product of the Io Plume Pele in the Discovery Picture
generated by Linda Morabito on 9 March, 1979. Limb fit to edge of
eruptive plume performed by Linda Morabito.
*Courtesy Linda Morabito*

Fred Peters, on one trip in, said:

"You can't base a find on one picture in Astronomy."
"Yes, but we have correlation with something on the surface," I retorted.
"On the surface!" Fred said, astonished.

I couldn't blame him, it was the initial psychological problem I had with the image; how resistant this discovery picture was to giving up its secret. The projection was huge, 0.15 of Io's radius, 250 kilometers above the limb of Io.

Fred said again that another observation would be needed for verification. I said, "It may not be there when we look again." I know Fred was stunned. I could only apologize to him for not saying more.

It was getting dark outside now, and most of the people in the Navigation area had gone home for the weekend. This was an incredible situation because no maps of Io existed yet to correlate Io's surface features with latitude and longitude; with one exception, however.

At the very moment we were working in the ONIPS room with the image of Io, a preliminary pictorial map of Io was being drawn by a U.S. Geological Survey artist upstairs in the Science Imaging Area, whom I would later come to know was P.M. Bridges. She was airbrushing in surface features of Io with breathtaking perfection in what looked like a partially completed photograph of a flat projection of Io, coming together on a latitude and longitude grid.

Peter warned me while we were still in the ONIPS room not to show any visible excitement if some of our theorized ejection centers turned up on top of what could apparently now be considered volcanoes. It was good advice. I remained very calm.

Peter and I had no choice but to function in the following way: every time we would derive a new ejection center we would jump up, run upstairs, and ask the artist if we could please look over her shoulder, or could she just look and see if she saw any kind of a surface feature at a certain latitude and longitude. Then, we would go back downstairs to try to use different parameters for our derivations, derive a new center, and do it all over again.

A small variation in our model of the cloud virtually whose crescent alone was visible, produced large variations in ejection centers, but none too far from the general region of the heart-shaped feature on Io. We repeated this process so many times based on varying models of the cloud that eventually the artist wanted to know where we were getting these locations.

Joe Veverka from Cornell, a Science Imaging Team member, one of the few who remained in this area now, whom I knew from Viking, watched Peter and I come and go with much suspicion. At one point, he put his hand up and came very close to stopping us, but he hesitated, and didn't.

As we sat back downstairs in the ONIPS room in between trips, Peter verbalized what I was feeling at the moment. "The more you work with it, the more you think it's real." Working with data in a new discovery puts the discoverer in a very unique situation for interpretation of that data. I always pay special, close attention now to what the discoverers of any new phenomenon have to say about their original studies; such as the discovery of the Martian meteorite ALH84001 holding evidence that magnetite crystals are of biological origin. I weight their original analysis very heavily, depending of course on what they did, in my opinion of the chances for ultimate validation of the discovery, in comparison to dissenting opinions that follow.

The ONIPS software was not able to conclusively tell us what we were looking for; it was not designed for this work, nor did it perform at any point well under crescent lighting.

Peter and I eventually abandoned that approach and began working geometrically with the image as in Figure 12. No model we tried got us far from the heart-shaped feature. When we had done what we could, Peter surprised me by telling me that Andy Collins had called Larry Soderblom, second in command of the Science Imaging Team earlier that afternoon to report the discovery. After a moment of consideration that didn't seem too bad. Obviously no one upstairs knew about the discovery, and Larry was at USGS in Flagstaff, Arizona.

Many years later, Andy Collins would be kind enough to share a little more insight into what did and did not happen that day. He had called Brad Smith in charge of the Science Imaging Team, who was back in Arizona, as he had called Larry Soderblom.

In retrospect, Andy had done exactly what he should. The Science Imaging Team was charged with this mandate. Discoveries on the missions were their directive. They knew about Stanton Peale's article, it could not have been overlooked as it was in the world of Voyager navigation engineering.

Andy's communication with Brad Smith failed that day. I understand he heard from Brad Smith the next morning who indicated he was aware that Andy had tried to let him know about something the day before. He asked Andy what it was that he had conveyed.

The hours and minutes of March 9, 1979, could have turned out more challenging than they even were by any turn of fate. But, it was a good thing for me there was no more than I had already handled. I called Andy at 8:15 PM after Peter left. Andy was at MTIS in the Space Flight Operations Facility. I was moments away from finding out I was simply too tired to even be able to form words of sentences anymore.

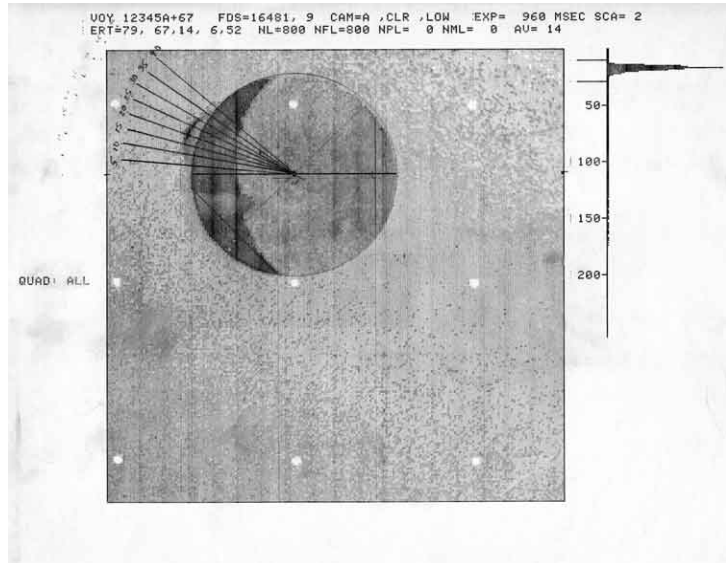

Figure 12. Original Optical Navigation Image Processing System data product of the Io Plume Pele in the Discovery Picture generated by Linda Morabito on 9 March, 1979.  Angles of eruptive plume drawn in by Peter Kupferman.
*Courtesy Linda Morabito*

I would later hear stories of how evidence of active volcanism on Io had been discarded in image processing techniques that had been specified by the Team to the IPL, the Image Processing Lab at JPL, but I have no concrete verification of that whatsoever. Dave Pieri, who was a PhD candidate at that time, and whose wife also worked for the Lab until he received his PhD, shared one story with me directly.

He said he spotted evidence of an eruption in a picture taken for science and had it explained away to him by a scientist on the Team.

What was crucial about making the discovery when I did, was that there was an immediate opportunity to go back to look at Io with Voyager 2 on its way in less than four months, and reprogram the mission on an Io intensive basis.

Had the discovery continued to elude the scientists, the opportunity would have been missed and not come again for another sixteen years when the Galileo spacecraft went into orbit around Jupiter.

Andy Collins called me as the hours went later into the night and invited me over to MTIS in the Spaceflight Operation Facility.  I did possess a red card that provided restricted access to the facility so that I was able to do that.  Andy showed me the hard-copy digifax of the discovery picture he was making.  He had done some stretches and high-pass filtering as well.

In the high pass filter product, I could see features on Io, the crescent was not completely overexposed, and a possible indication of brightening in the region of the cloud, indicating that the best model of its ejection center might be on the front of the disk coming at us, rather than originating from behind.

I was very candid with Andy about my concerns about the discovery, but mostly I talked about what the science meant to me. I was so exhausted now the words sometimes didn't form correctly.

I told Andy this discovery was now part of me and could never stop here for me. In ways I could not have predicted then, that was exactly correct. I would indeed go on to head another discovery on another planet Mars accomplished by another generation for The Planetary Society on a huge, international project.

Andy seemed a little speechless. He may have been able to tell how hard I was fighting to conceal what was past exhaustion now. In my mind, since I could come up with no further thinking of any kind, my ability to hang on to the image was threatened.

Andy volunteered the fact that he had called Larry Soderblom in Arizona. I was glad he mentioned it. Depending upon whether Larry took the train in or flew back, we would meet either Sunday night or Monday morning, OPNAVERs and scientists in a spirit of complete cooperation.

Andy asked if he could call me at home over the weekend to let me know whether the meeting would be Sunday or Monday. Andy was being very, very kind. As he had surely known when he had asked me, he had my permission to call me at home.

I would receive another call at home in the coming week. The head of the Science Imaging Team, Brad Smith called one morning to ask what was my background. I am guessing he had been asked by the press often, and had no answer. When I told him I was an astronomer, I believe I heard an audible sigh of relief.

*National Geographic*, *Smithsonian*, *Scientific American*, and *Time Magazine*, all covered my discovery. *Time* noted in its January 21, 1980 issue, that the volcanic activity on Io was the cover story on all the three aforementioned magazines in their January 1980 issues, total circulation 12,750,000. Rather than having the same volcanic eruption pictured on all three magazines be a nightmare, instead the magazines viewed this as confirming their good judgment, the story in *Time* goes.

Andy's phone call woke me up at 9:00 AM. He told me that the meeting with the scientists from the Science Imaging Team would take place Monday morning at 9:30 AM. I decided I could generate some impressive computer products from ONIPS relating to the image to re-enforce the way this find had come to light. I kept thinking what I would say to the Imaging Team during that meeting. I kept thinking of Jocelyn Bell and the history surrounding her discovery. If they wanted to hear that I knew that history, I would tell them.

I went to the ONIPS area Saturday afternoon to generate the computer products that I had thought to do. I dumped the entire image of the satellite and the anomaly, cut and pasted it together and was tracing out Io's limb when the door opened behind me. The entire Navigation area was deserted when I arrived, and I couldn't imagine who it would be. It was Ed Stone, the Voyager Project Scientist. He asked me if he could bring Bob Parks down. Parks was the Voyager Project Manager. I said yes.

I did not introduce myself when Parks arrived, for despite their presence I was still so completely engrossed in the phenomenon itself, I could hardly contain myself in wanting to talk about it and share it.

I took off like a shot, displaying pictures on my monitor.  All I wanted to do was share with them the wonder of what I was seeing.  I wanted to know if they knew the cloud was over the heart-shaped feature and had they seen the digifax?  No, they didn't know that.  Their eyes glistened.  I mean that literally.

I spoke technically and competently.  But, not introducing myself was not a transgression of assertiveness in my estimation.  It seemed natural and correct.  The significance of the discovery far overshadowed me, and I did not feel I was cheating myself by allowing that statement to be made.  They obviously knew what was done here, or they would not have come to see me.

They came to ONIPS, a viable place for this discovery to have been made.

Ed Stone volunteered they had commanded the spacecraft at high risk to turn on the PPS (Photopolarimeter) instrument and look back for traces of the cloud.  It had apparently been decided just earlier based on Ed Stone's exposure to my discovery the day before.  Ed Stone did not have to tell me that, but he did.

"If it's verified," Ed Stone went on to say, "it will be a marvelous discovery." At the time, I had taken that as a general statement.  They went along their way and I continued where I had when they arrived. Much later in the evening, I suddenly stopped what I was doing. I remembered Ed Stone was looking me straight in the eye when he said that, and saying that to me.

It's Verified!
At 9:15 AM Monday morning, March 12, 1979, Steve and I were preparing for the 9:30 AM meeting with the scientists.  The Navigation area was full.  The phone rang. It was Peter Kupferman.  "You'd better get up here!" he exclaimed.  "They've found volcanoes all over the place!"

I could hear people actually screaming in the background.  "Everybody's gone crazy!" Peter cried.  Steve and I marched out of the ONIPS room to the Navigation area.  I had computer outputs  grasped tightly in my hand.

I yelled back at the Navigation Team, "It's verified!"  That was a good feeling.  A cheer of elation went up.

When we arrived upstairs, rather than a meeting it was Pandemonium, the area filled with celebrating scientists pouring over pictures.

I could see Brad Smith, the head of the Imaging Team, standing in a corner of the crowded room, looking up at an unenhanced science pre-encounter image of a volcanic eruption displayed on a monitor.  Large as life.  But, no one had noticed it.  Brad Smith looked detached from everything.

When we came into the room, he did not look directly at me.  He said, as if talking to the air, "Linda noticed it first."  I thought of Andy and a type of work he had to do that may indeed have been very much a part of what he did so well throughout the mission that had less to do with the technical issues, but rather human interactions.  I knew I had survived.

In addition to bringing the discovery forward to fruition, I wanted to be able to continue in my field in the only way I knew that I could, without the bitterness about what people can do to one another as I had learned from history.  Larry Soderblom wasn't having any problem at all.  He said, "Don't leave town," and actually looked at me when he said it.

Frank Jordan aptly made the point later that Navigation should get a publication in the *Science* issue that presented the results of the Voyager 1 encounter with Jupiter to the world. L.A. Morabito, S.P. Synnott, P.N. Kupferman, and Stewart A. Collins, in the order of our involvement published "Discovery of Currently Active Extraterrestrial Volcanism" in the 1 June 1979 issue of *Science* **204**.

By then, a second volcanic eruption, which I had not even considered, was subsequently noticed on the terminator in the discovery picture, the boundary between day and night on Io.  At the moment the picture was taken, it was projecting above the dark surface into the sunlight.  The volcano of this eruption was eventually named Loki.  I had discovered the volcanic plume of the volcano that would be called Pele, both Hawaiian Gods.

My eleven page letter which describes every moment of the discovery dated March 15, 1979, written to and because of Gibson Reaves, who passed away in May of 2005, is signed "Your former student, friend, and admirer, Linda Morabito."

4. The Future of Science

Io, although magnificent, is not unique.  There are several moons of the outer solar system that show signs of liquid water or an ocean, some similar to Io in that their internal heat can be attributed in good part to tidal heating (Solomonidou 2011).  Humanity is now in a situation where they may see a future of habitable world discoveries in the exomoons of the galaxy and likely the universe; millions if not billions or trillions of such worlds in our galaxy alone.  My find of the active volcanism on Io was significant. It defined a new class of worlds that derive their internal geologic activity from an external circumstance that causes a tidal bulge to flex their surfaces.

Many of the key ingredients history has shown us are a part of discovery were manifested in those many hours after it was believed the Voyager 1 mission was over in terms of all the wonders in the realm of Jupiter having been revealed.  Resistance to new ideas in science, dismissing of observations which do not fit into expectation, not giving up for fear of one's circumstances appearing hopeless, are among many the characteristics of what I saw displayed, and chose to do when my discovery arose.

I was perhaps born aware of my love of astronomy and science, but I also had a lifelong mentor play a significant role in my success in science.  He taught me the history of science, which likely permitted me to play a role in it. Our educators, our society, we, must commit to establishing an educational system that permits young people to reach their potentials and participate in real scientific research while still formulating their direction in life.  We must tell them that science and exploration is important!   If we teach them about the past and provide the priorities they will in earnest emulate, they will assure the future of science.